\documentclass[12pt,a4paper]{article}
\pdfoutput=1
\usepackage[utf8]{inputenc}
\usepackage[T2A,T1]{fontenc}
\usepackage[english]{babel}
\usepackage{amssymb,amsfonts,amsmath,mathtext,cite,enumerate,float} %
\usepackage[dvips,pdftex]{graphicx}
\usepackage{geometry} %
\usepackage{ifpdf}
\usepackage{hyperref}
\title{Simulation of an axial vircator with a three-cavity resonator}

\author{P.V. Molchanov, E.A. Gurnevich, V.~V.~Tikhomirov\thanks{E-mail:vvtikh@mail.ru}, S.E. Siahlo}

\begin{document}
         \maketitle
         \begin{center}
                  Research Institute for Nuclear Problems, Belarusian State University,\\
Bobruiskaya 11, 220030 Minsk, Belarus
         \end{center}

\begin{abstract}
We simulated an axial vircator with a three-cavity resonator and expected generation efficiency 6--7$\%$. For adequate description of physical processes taking place inside a vircator we used two independent PIC codes: self-developed INPIC and free XOOPIC. Based on both the analysis of the vircator proposed in [1] and consideration of the devices operating at cathode-anode voltages under 450 kV we suggest 3 possible designs of a three-cavity resonator such that enable one to produce High Power Microwave in GW power range.
\end{abstract}

\section{Introduction}

Virtual cathode oscillators (vircators) demonstrate significant
advantages over other microwave sources: ability to operate
without a guiding magnetic field, rather short operation region,
enhanced tunability of the operation frequency, high output power,
and relative simplicity of resonator design.  However, low
efficiency and instability of the generation  frequency are
typical disadvantages of most vircators.

The efficiency of vircators can be increased, particularly,
by setting resonant conditions for the beam energy--to--HPM (High Power Microwave) conversion. The authors of \cite{1} have described the design of an  axial vircator  capable of high power efficiency (6--7$\%$). According to \cite{1}, the electron beam produced in a 700 kV and $\sim$ 24 kA diode and interacting with a multicavity
resonator at a frequency of about 4 GHz provided the average
output power of about 1 GW.

\begin{figure}[h]
       \centering
        \includegraphics[width=0.7\linewidth]{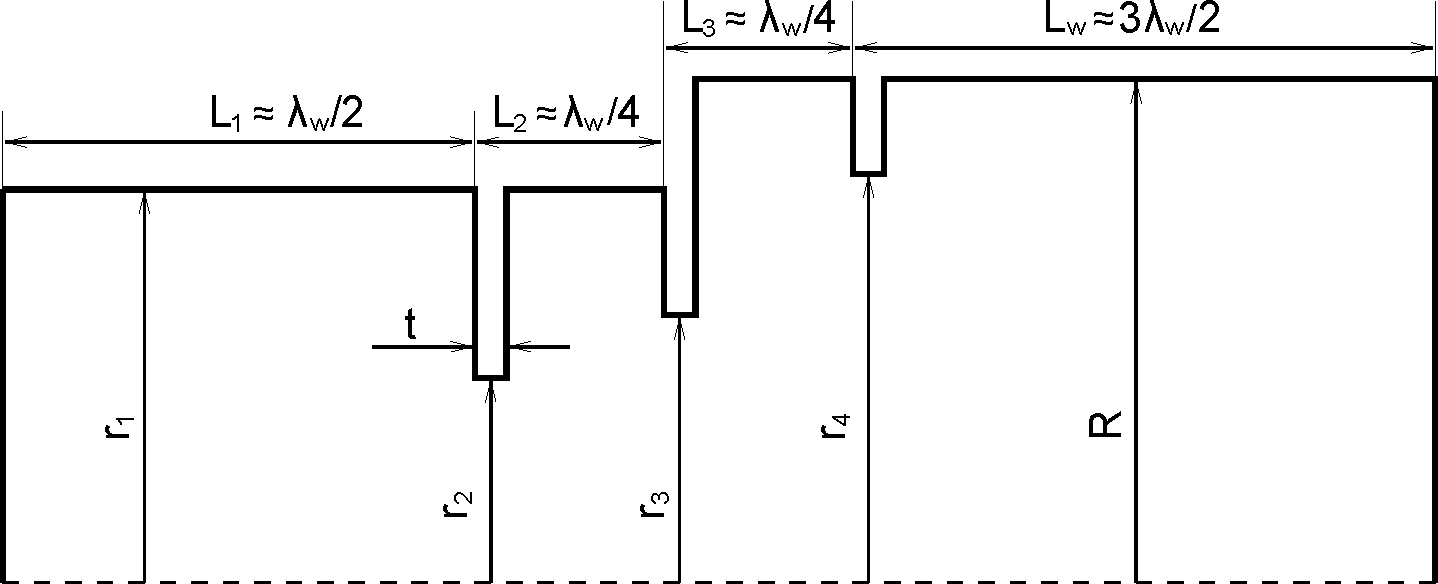}
              \caption{Geometry of a three-cavity resonator.}
    \end{figure}

The aim of the present work is to develop such geometry of a
multicavity resonator for an axial vircator  that could provide HPM
generation at a frequency from 3 to 4 GHz with the efficiency
higher than 5$\%$ at a cathode voltage less than that used
by the authors of \cite{1}.

Various configurations of  vircators were
simulated with the self-developed code \cite{2}, which we
called INPIC (Institute for Nuclear Problems Particle In Cell
code). Like any other PIC code, INPIC combines finite-difference
formulation of  Maxwell's equations and the finite-size particle
method. As is known  \cite{3}, use of discrete (or
finite-difference) methods for the solution of Maxwell's equations
implies regular corrections and smoothing (numerical filtering) of
numerical solutions for fields, charge and current densities.
The techniques and methods used for these correction procedures involve a number of parameters which can be chosen from comparison with other simulation or experimental results rather than from theoretical consideration. To gain both validation and experience in fixing these parameters we will extensively use here simulation results obtained using a free XOOPIC
code. The main difference between simulations with INPIC and
XOOPIC \cite{4} lies in the simulation procedures for the electron
injection from the cathode and in the processes occuring in the
cathode-anode zone.

With  INPIC, the charge is injected in a plane that is one quarter
cell width away from the emitting surface of the cathode.  The
value of the injected charge is governed by the Gauss theorem for
cells adjacent to the surface and by the electric field vanishing
on the emitting surface. In other words, this means that the value of the
electron-beam current is not an initially preset parameter, but
is calculated from the physical laws at each time step.

\begin{figure}[h]
       \centering
        \includegraphics[width=0.6\linewidth]{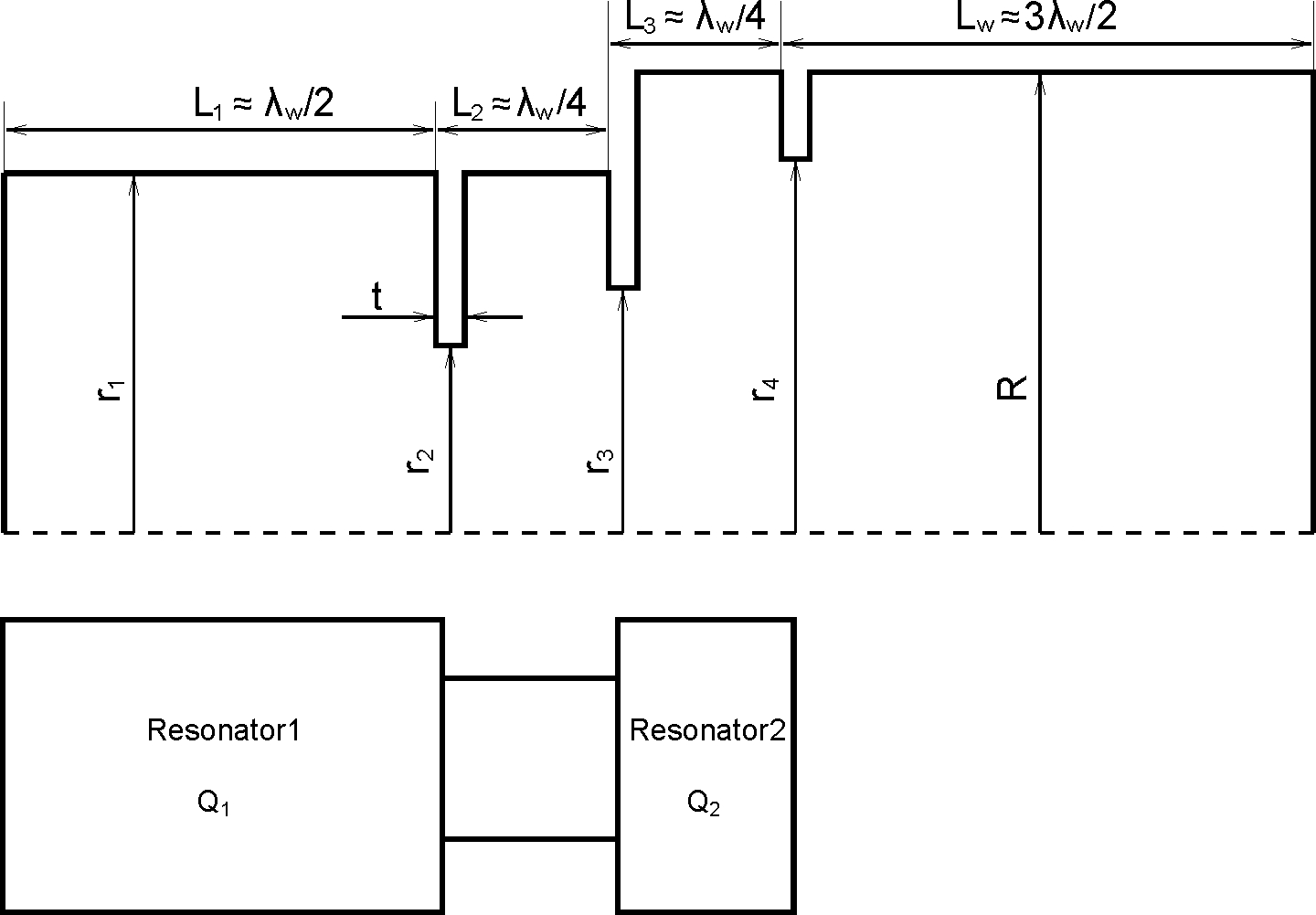}
              \caption{Structure and equivalent schemes of the three-cavity resonator.}
    \end{figure}
Simulations performed with XOOPIC do not consider the processes in
the cathode-anode gap: the electron beam having a fixed electron
current and energy is injected directly into the  first resonator
cavity. The electrons reflected from the virtual cathode are left
out of the consideration if they  transmit the injection plane
(anode mesh).

The paper is arranged as follows. The second section reconstructs
the dimensions of the system  \cite{1} from the graphs given therein
for simulated and experimental radiation spectra and power at a specified
diode voltage and current. The third section
describes the procedure used to select the geometry of a
three-cavity  resonator for generating radiation in the frequency
range from 3 to 4 GHz at the electron beam energies from 300 to
450 keV (XOOPIC) and diode voltage of 450 kV(INPIC) and
gives several variants of the selected geometries. We simulate the radiation
spectrum and output power of the axial vircator  using both INPIC and XOOPIC
for the suggested geometries of  three-cavity resonators
and it is shown  that the power efficiency can be as
large as  about 5$\%$.

\section{Simulation of a three-cavity resonator \cite{1}}

Fig.1 shows the schematic geometry of  the three-cavity
resonator under study. The authors of \cite{1} specify only some
dimensions of the structure: the anode diameter equal to the
diameter of the first section $d_1=2r_1=80$mm; the cathode
diameter -- 64 mm; the dimensions of the cathode-anode gap -- 14mm
at which the maximum efficiency  was obtained, and the geometric
transparency of the anode mesh -- 70 $\%$; they also give the
frequencies of the axial vircator  that correspond to different
values of the cathode voltage: 4.1 GHz at 630 kV and  3.98 GHz at
700 kV, respectively.

All other dimensions in Fig.1 were reconstructed using the
POISSON SUPERFISH code. In simulating with POISSON SUPERFISH, we
choose all dimensions so as to provide the coincidence between the
calculated resonant frequencies  and those given in \cite{1} (4.04
and 4.16 GHz). Thus obtained data are listed in Table 1.

\begin{figure}[h]
       \centering
        \includegraphics[width=0.4\linewidth]{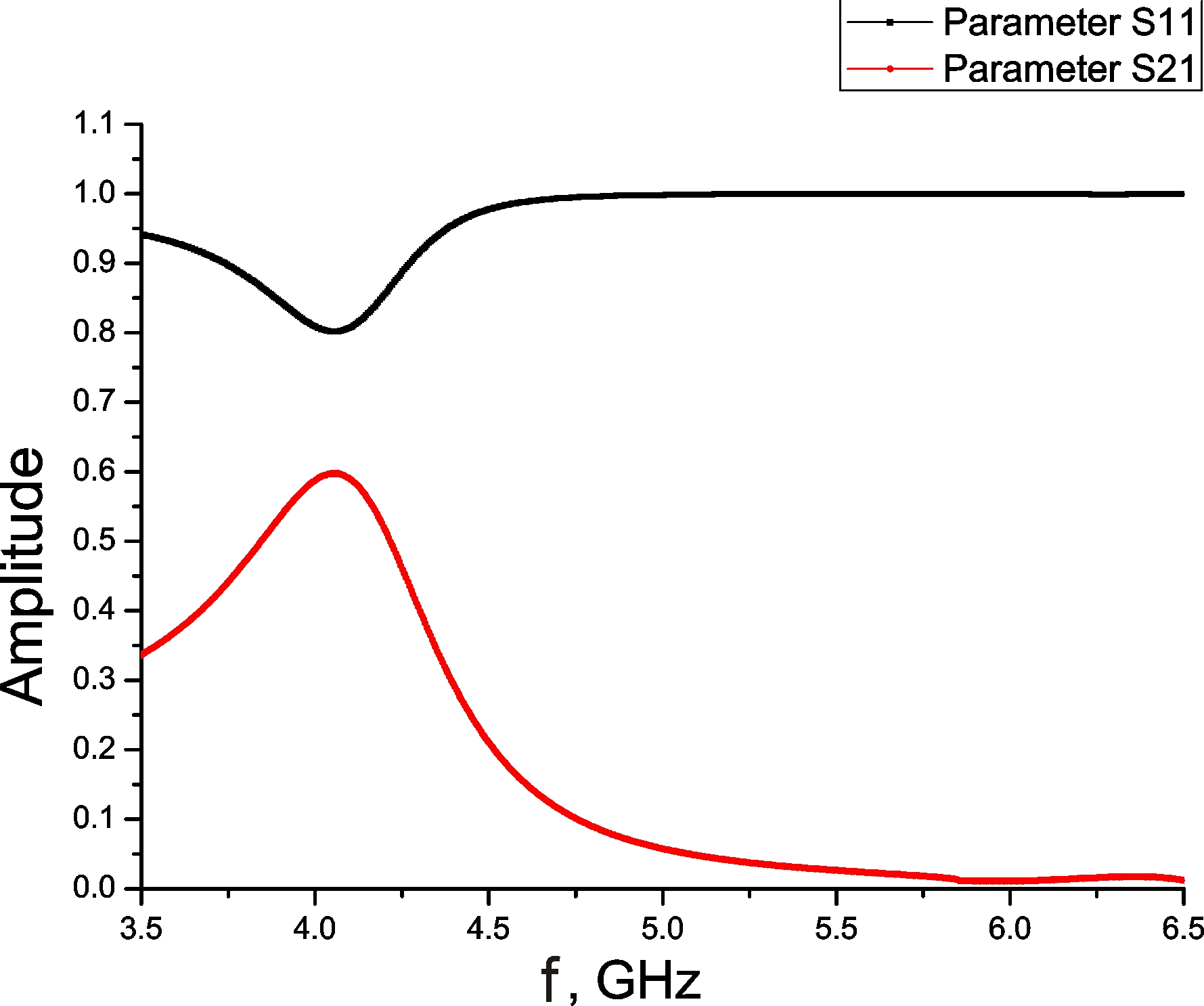}
              \caption{Simulated S parameters of the three-cavity resonator.}
    \end{figure}

The total length of the resonator appears to be 261 mm. According
to the simulation results reported in  \cite{1}, the average power of 1 GW
was achieved at diode voltage and current of 630 kV and 24
kA, respectively, while at diode voltage of 700 kV, the average
power was 1.8 GW with the instantaneous power peaks
greater than 4 GW.

\medskip
\mbox{Table1}
\medskip

\begin{tabular}{|l|l|l|l|l|l|l|l|l|l|l|}
  \hline
  Parameters of the resonator (Fig.1) & $L_1$ & $L_2$ & $L_3$ & $L_w$ & $r_1$ & $r_2$ & $r_3$ & $r_4$ & $R$ & $t$ \\
  \hline
  Dimensions, mm & 54 & 24 & 26 & 157 & 40 & 20 & 33 & 37 & 45 & 3 \\
  \hline
 \end{tabular}
\bigskip

In the first section of the resonator the main energy exchange
between the electron beam and the electromagnetic field takes
place. The dimensions of this section (radius $r_1$ and length
$L_1$) relate to the operation frequency $f_{01}$  through the
condition of resonance $TM_{011}$  of a cylindrical resonator

\begin{equation}
\label{eq1}
f_{01}=\frac{c}{2\pi} \sqrt{\frac{\nu_{01}^2}{r_1^2}+\frac{\pi^2}{L_1^2}},
\end{equation}

\noindent where  $c$  is the speed of light, $\nu_{01}=2.4048$  is
the first zero of the Bessel function $J_0$.  One should bear in
mind that as the electron beam is injected, the frequencies of the
resonator's eigenmodes change slightly (under the conditions
considered here the frequency $f_{01}$ is reduced by 100 -- 200
MHz).

Two shorter sections (cavities) designed for
effective matching of the resonant section with the output
waveguide. Their functionality is understandable
from the analysis of electrodynamical properties of the
three-cavity resonator (Fig.1) which can be considered as a
band-pass microwave filter  composed of two resonator sections
with a quarter-wave cylindrical insert between them \cite{5}
(Fig.2).

The length $L_3$ determines the magnitude of the reflected wave in
the filter pass band; which has a minimum at $L_3\sim \lambda _w/2$.
To ensure sufficient reflection of a wave from the resonator's
boundary which provides the internal feedback at vircator oscillation
frequency we choose $L_3\approx \lambda _w/4 $. For the indicated
dimensions, the filter pass band has a maximum corresponding to
the generation frequency of the vircator. We show in Fig.3 the transmission and reflection coefficients of the electromagnetic wave  (parameters S21 and  S11) for $TM_{01}$ wave in the
resonator.

Using the parameters in Table 1, we simulated the three-cavity
axial vircator (Fig.1) with  XOOPIC.  The diameter of the injected
beam was taken to be equal to that of the cathode -- 64 mm,  and
the beam was assumed to be solid and homogeneous. The energy of
electrons was set to be 630 and  700 keV. The injected current was varied so as to provide the best possible coincidence between the simulated radiation power and spectrum obtained in this paper and in \cite{1}.

\begin{figure}[h]
       \centering
        \includegraphics[width=0.98\linewidth]{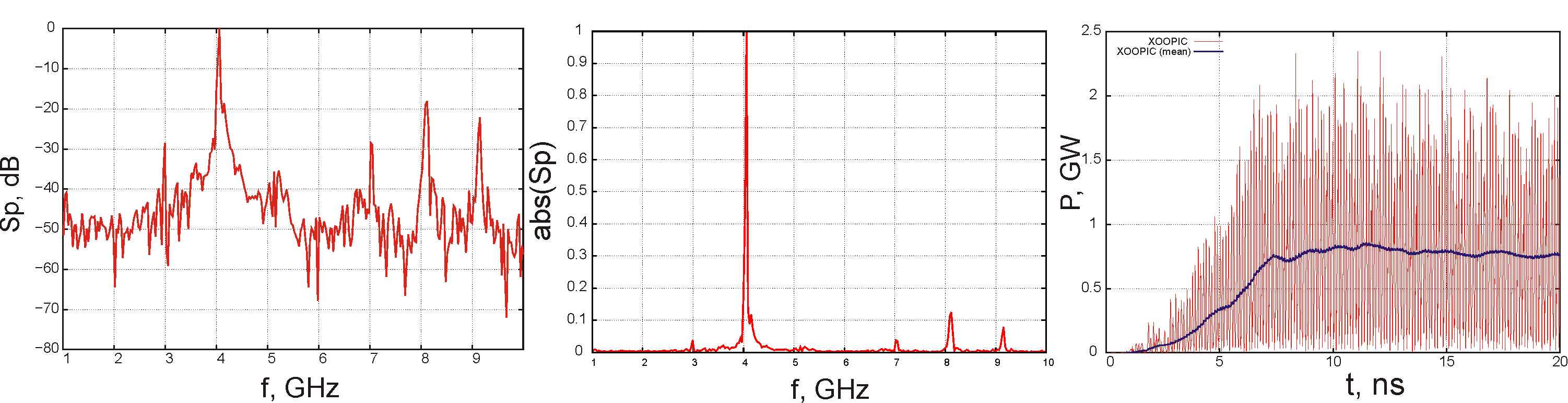}
              \caption{Radiation spectrum (left and center) and power (right) in the three-cavity axial vircator (Fig.1, Table 1) simulated with  XOOPIC for the injected beam of energy 630 keV and current 21 kA; Blue curve - averaged (with a period of 1 ns) value of the output power.}
    \end{figure}
\noindent

\begin{figure}[h]
       \centering
        \includegraphics[width=1.\linewidth]{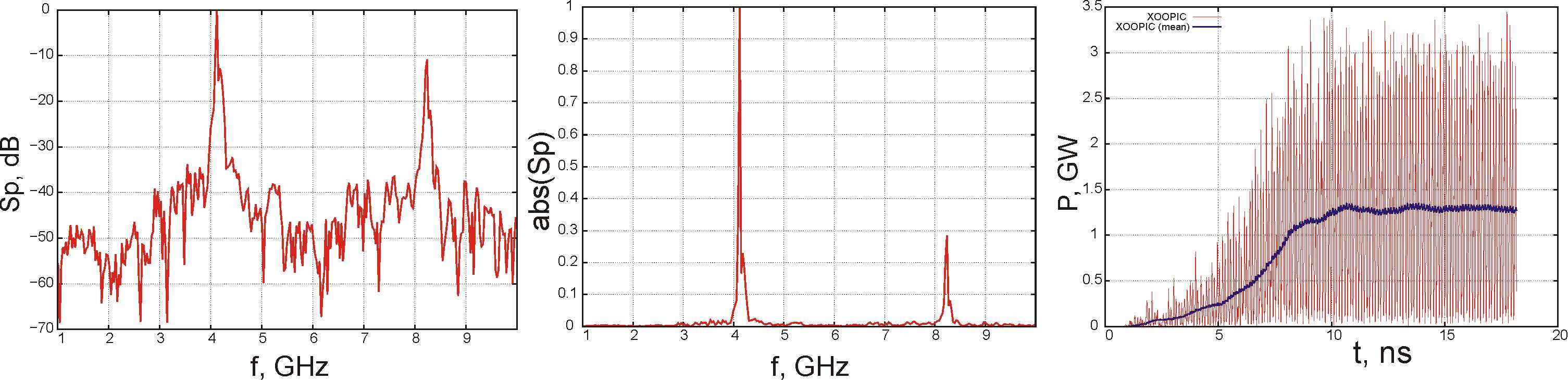}
               \caption{Radiation spectrum in the vircator (Fig.1, Table 1)
               simulated with  XOOPIC for the injected beam of energy 700 keV and current 24 kA.}

    \end{figure}
\noindent

Fig. 4 shows the simulation results obtained with XOOPIC
for the case of  injection of a 630 keV,  21 kA  beam.  The
radiation frequency of  $\sim$ 4.1 GHz  and efficiency of about
6.5$\%$  agree well with the results reported in \cite{1} for the
diode voltage of  630 kV, but   the used values of the  injected
current  and the obtained  values of the average output power were
less than those reported in \cite{1}, which may result from
the  incomplete coincidence of the parameters used in
simulations. Simulation of  the axial vircator with the  electron beam energy of
700 keV and  current of  24  kA (Fig.5) gives greater values of frequency (4.15 GHz versus 3.98 GHz)
and slightly less values  of  the instantaneous and average  power than those reported in \cite{1}.

\begin{figure}[h]
       \centering
        \includegraphics[width=0.8\linewidth]{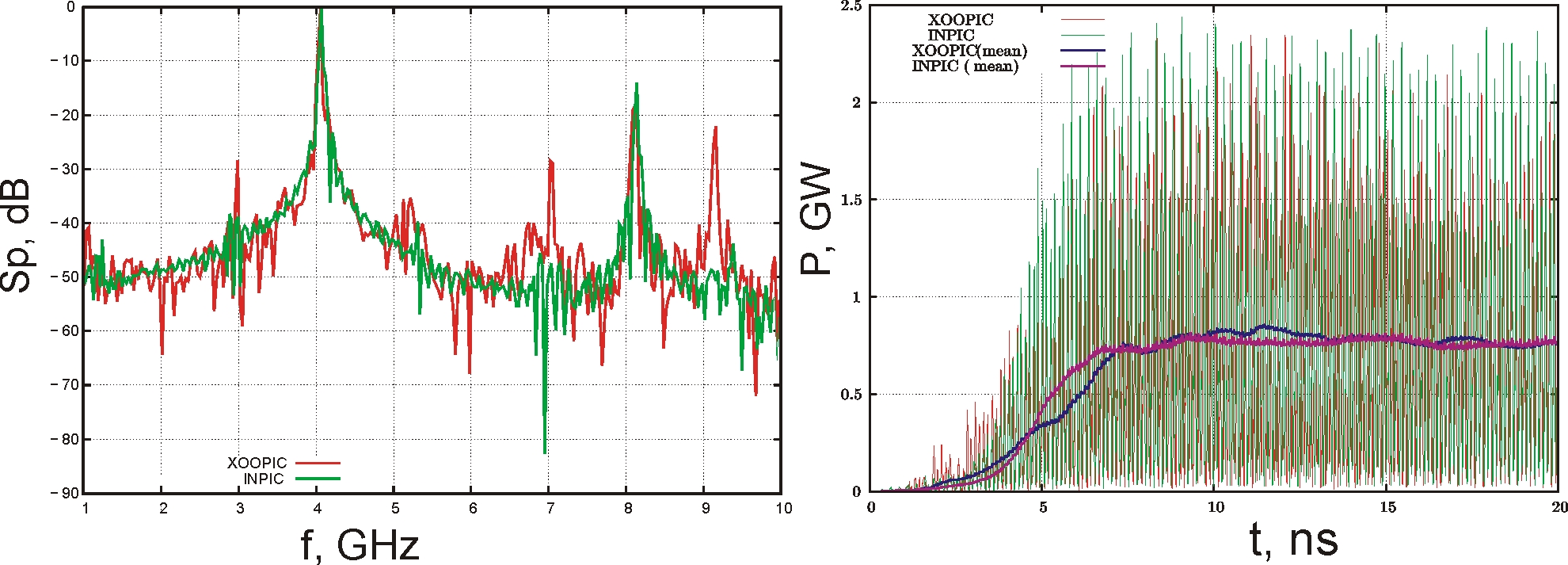}
               \caption{Simulation results obtained with INPIC at cathode voltage of 630 kV
               vs those obtained with XOOPIC for the beam energy of 630 keV spectrum (left), output power (right).}
    \end{figure}
\noindent

   \begin{figure}[h]
       \centering
        \includegraphics[width=0.8\linewidth]{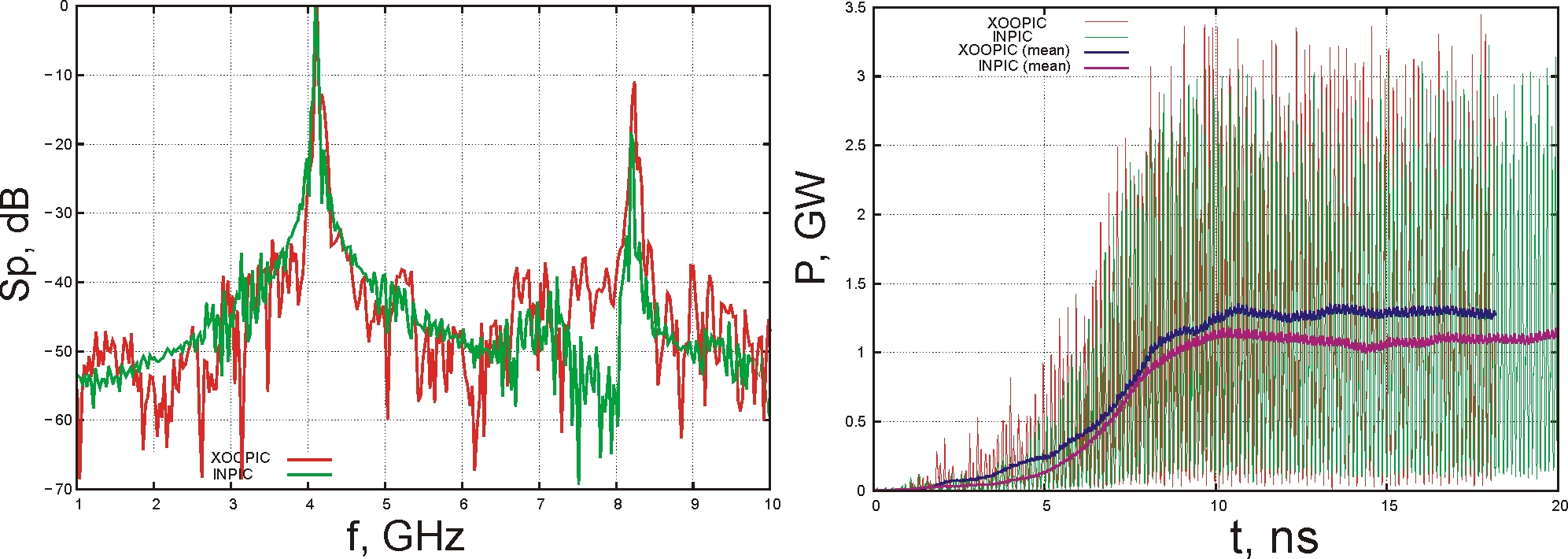}
              \caption{Simulation results obtained with INPIC at cathode voltage of 700 kV
              vs those obtained with XOOPIC for the beam energy of 700 keV spectrum (left), output power (right).}
    \end{figure}

\noindent

To reproduce the radiation spectrum and the output power reported
in \cite{1}, when simulating with INPIC of the axial vircator with
the same geometry of the three-cavity resonator (Fig.1, Table 1),
we  set the cathode to be ring-type. Because the value of the charge injected
from the cathode was calculated from the Gauss theorem for each
time step separately in each cell, it varied over the
cells, as well as over the time steps. As the electrons propagate
from the cathode to the anode mesh they are deflected from the initial
direction  by the eigenfield of the beam and weak nonuniformities
of the accelerating field, which are determined by the shape and
the finite size of the cathode, and so the beam injected into
the resonator is neither  continuous nor homogeneous, in contrast to
the  simulation using XOOPIC. The
selection criterion for the cathode parameters and the
cathode-anode gap was a fit with \cite{1} in both the output
power and the radiation spectrum. For the voltage in the
cathode-anode gap equal to 630 kV, the outer and inner diameters
of the cathode are selected to be 63 and 21 mm, respectively. The
current injected into the resonator (taking into account the anode mesh
transparency) is about 21 kA for a 14 mm cathode-anode gap. When
modeling the vircator whose voltage is 700 kV, the outer and
inner diameters of the cathode are set 61 mm and 19 mm,
respectively, the cathode-anode gap is selected to be 14 mm.

\begin{figure}[h]
       \centering
        \includegraphics[width=0.9\linewidth]{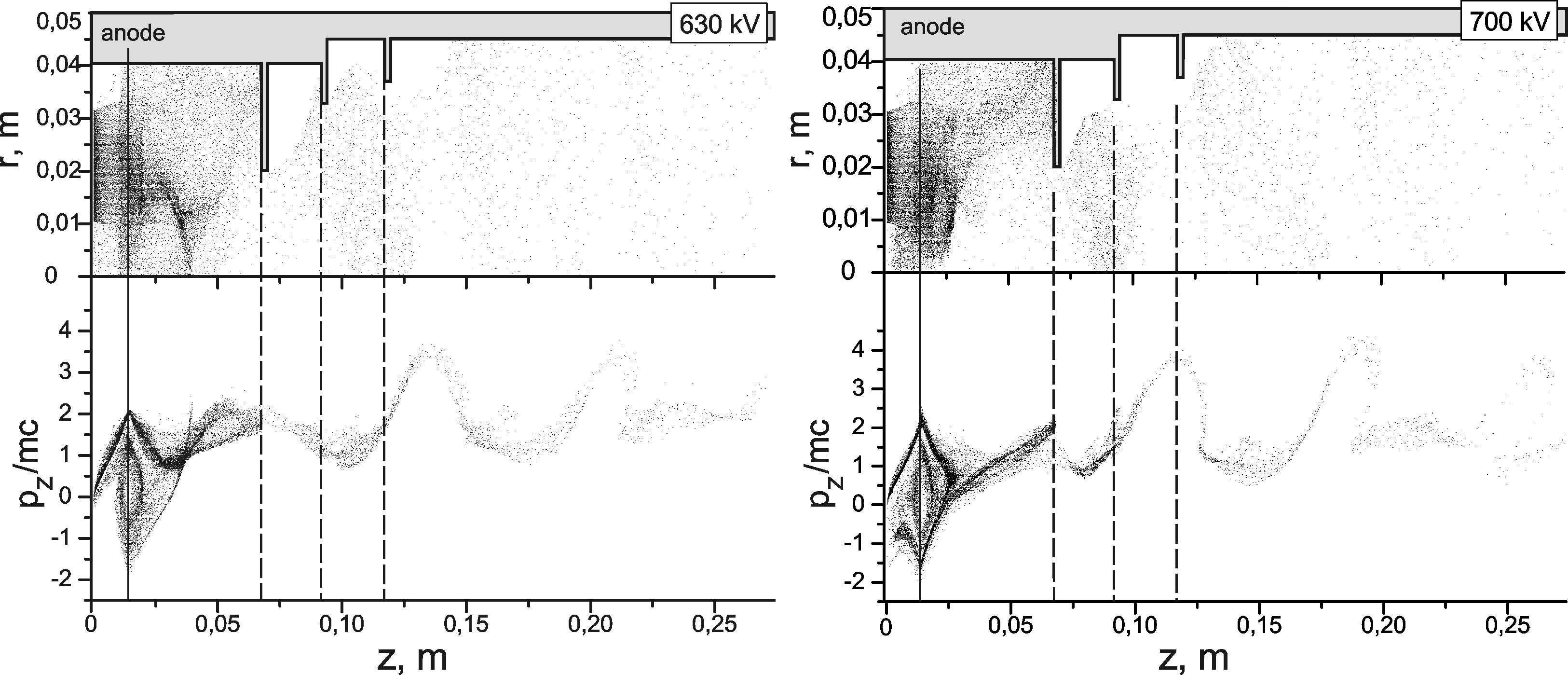}
               \caption{Configuration (top) and longitudinal (bottom) phase
               portraits of the beam in the three-cavity axial vircator for cathode voltages of 630 (left) and 700 kV (right).}
    \end{figure}

The comparison of  the simulation results obtained with INPIC and XOOPIC
(Figs 6 and 7) demonstrates a good fit of both the radiation spectra and the output power. The simulation results given here adequately describe the behavior of the system in \cite{1} for
the electron beam parameters  specified therein.

\begin{figure}[h]
       \centering
        \includegraphics[width=1\linewidth]{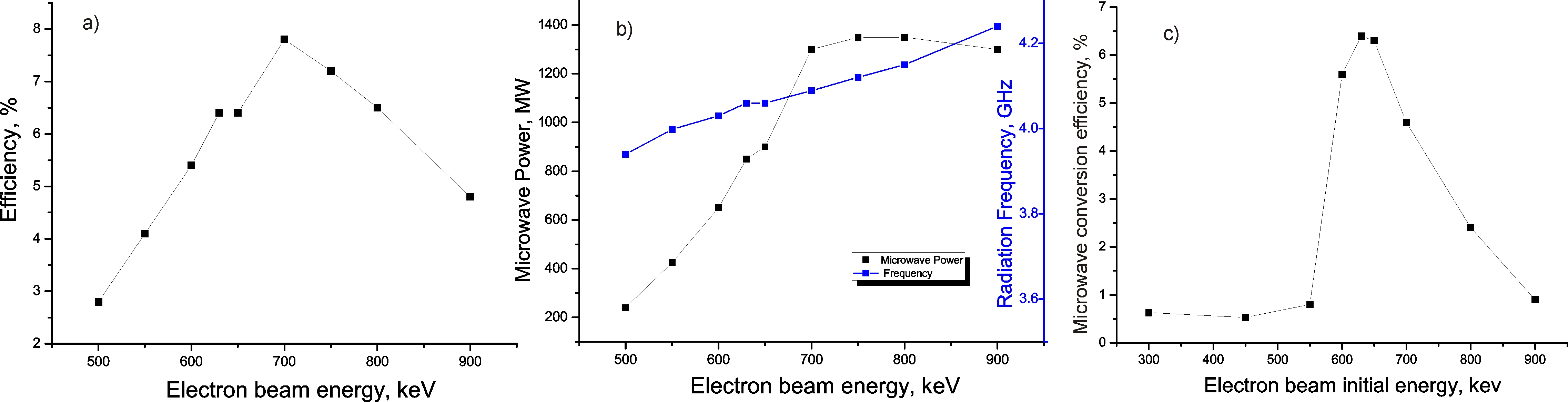}
               \caption{XOOPIC simulation results for generation efficiency (a), output power and generation frequency (b) and generation efficiency (c) vs the electron-beam energy at constant impedance of the beam.}
    \end{figure}

It was demonstrated in \cite{2} that in the region behind the
virtual cathode in a three-cavity resonator of an axial vircator,
the electrons can accelerate up to the energies
$\sqrt{p_z^2c^2+m^2c^4}-mc^2$, which are more than twice as large
as the maximum acceleration energy of electrons in a cathode-anode
gap (1.24 in the first and 2.84 in other sections). Indeed, as
is clearly seen in phase portraits (Fig. 8), in the first
resonator cavity, most of the high-energy particles rapidly
reach the resonator's walls and  become retarded by the first
annulus. In this way high-energy particles are removed from the
acceleration process, and the energy of the electromagnetic field
that was not expended into worthless acceleration of particles
will contributes to the increase of the output power.

\begin{figure}[h]
       \centering
        \includegraphics[width=0.8\linewidth]{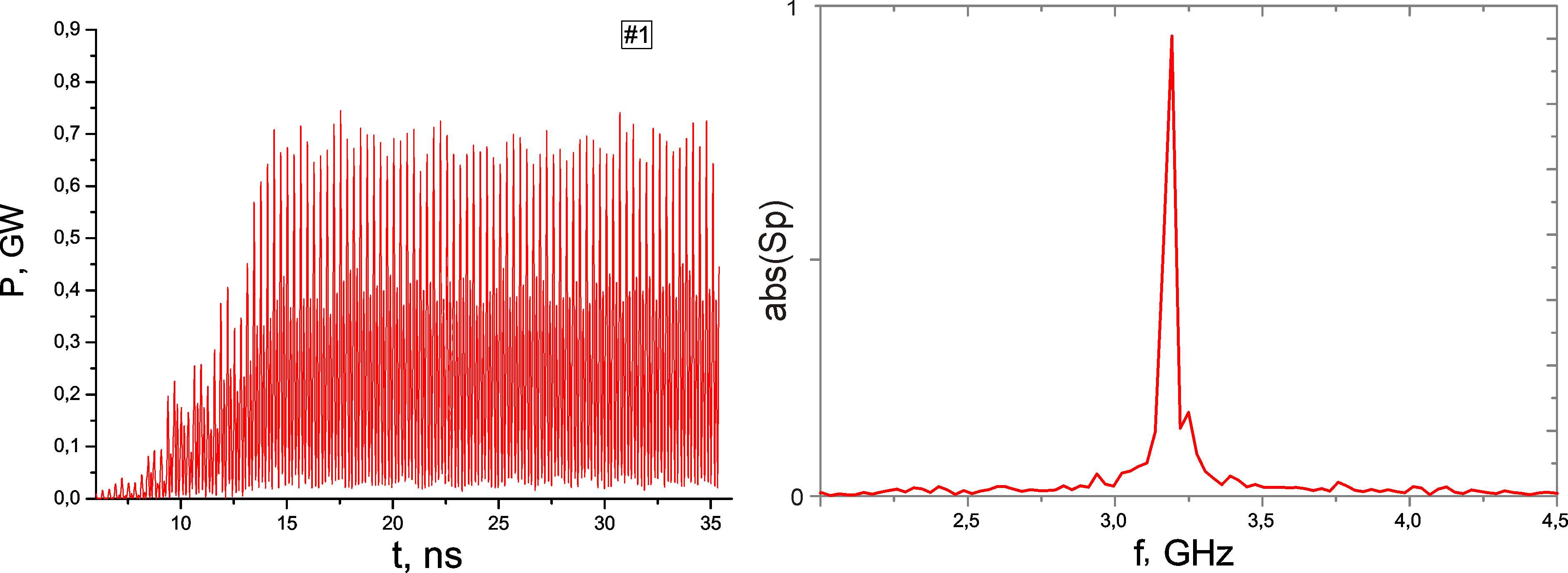}
              \caption{Output power and spectrum for the axial vircator with a three-cavity resonator $\bf{\sharp}$1 simulated with the INPIC.}
   \end{figure}

\section{Suggested geometries of a three-cavity resonator for a 300-450 keV  electron beam (XOOPIC) and 450 kV diode voltage (INPIC)}

Let us consider the possibility of effective HPM generation in the
given resonator geometry  \cite{1} for the electron energy less
than 600 keV. Studying how the  output power and the generation
efficiency in such systems depend on the electron-beam energy at
constant impedance of the beam, we found a noticieble maximum in
the range from 600 to 750 keV (Fig. 9). The maximum output power
is also achieved at 700 keV, dropping sharply as the beam energy
is decreased. An increase in the electron energy beyond 750 keV is followed by a loss of generation regime stability and does not provide a rise in the output power. The analysis given in the previous sections shows that  effective generation at electron energies less than 600 keV is achievable only with a three-cavity resonator having the dimensions different from those given in \cite{1}.

   \noindent

  \begin{figure}[h]
       \centering
        \includegraphics[width=0.5\linewidth]{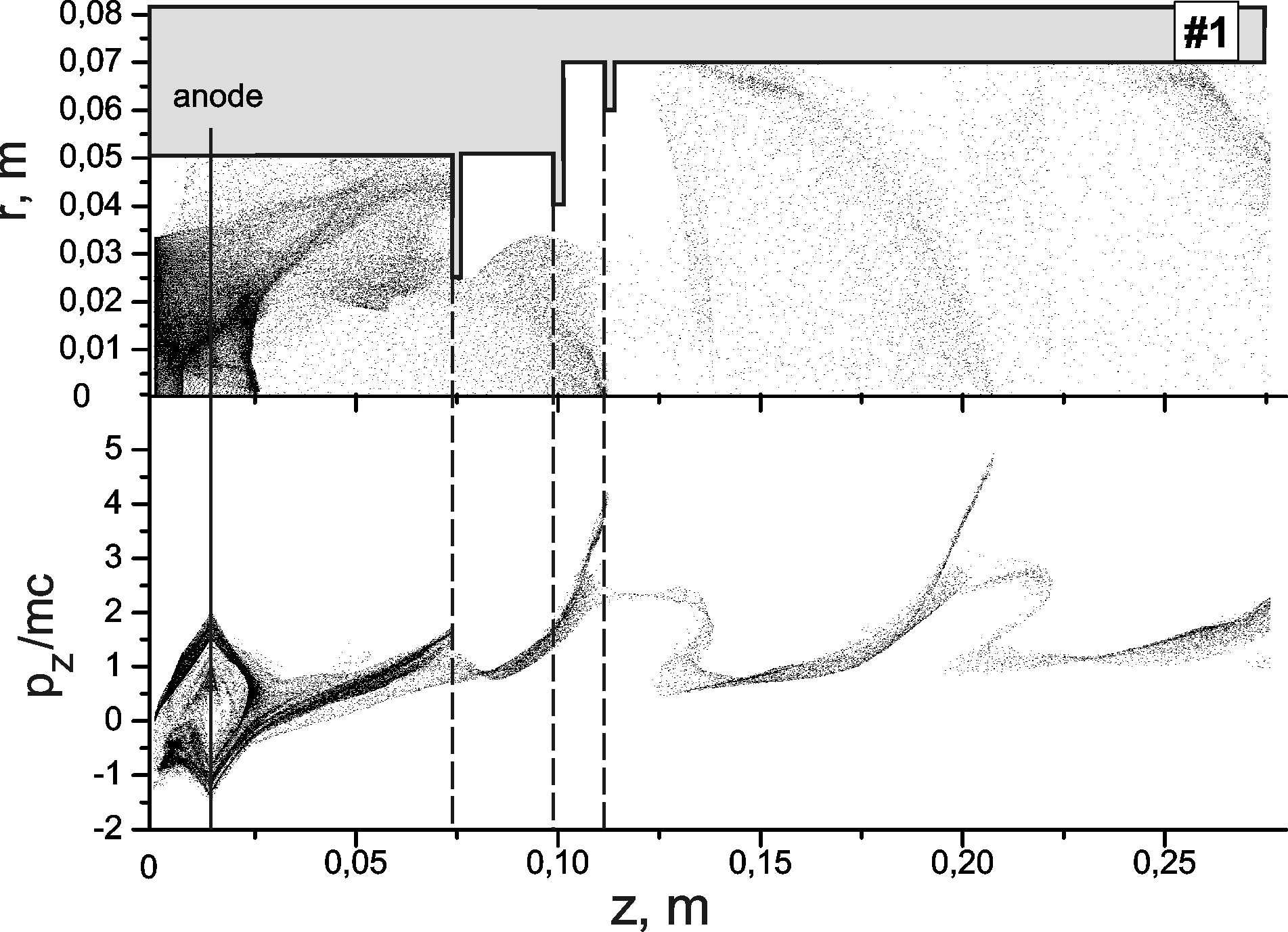}
              \caption{Configuration (top) and longitudinal (bottom) phase
               portraits of the beam in axial vircator with  three-cavity resonator $\bf{\sharp}$1 simulated with the INPIC.}
 \end{figure}

   \noindent

  \begin{figure}[h]
       \centering
        \includegraphics[width=0.8\linewidth]{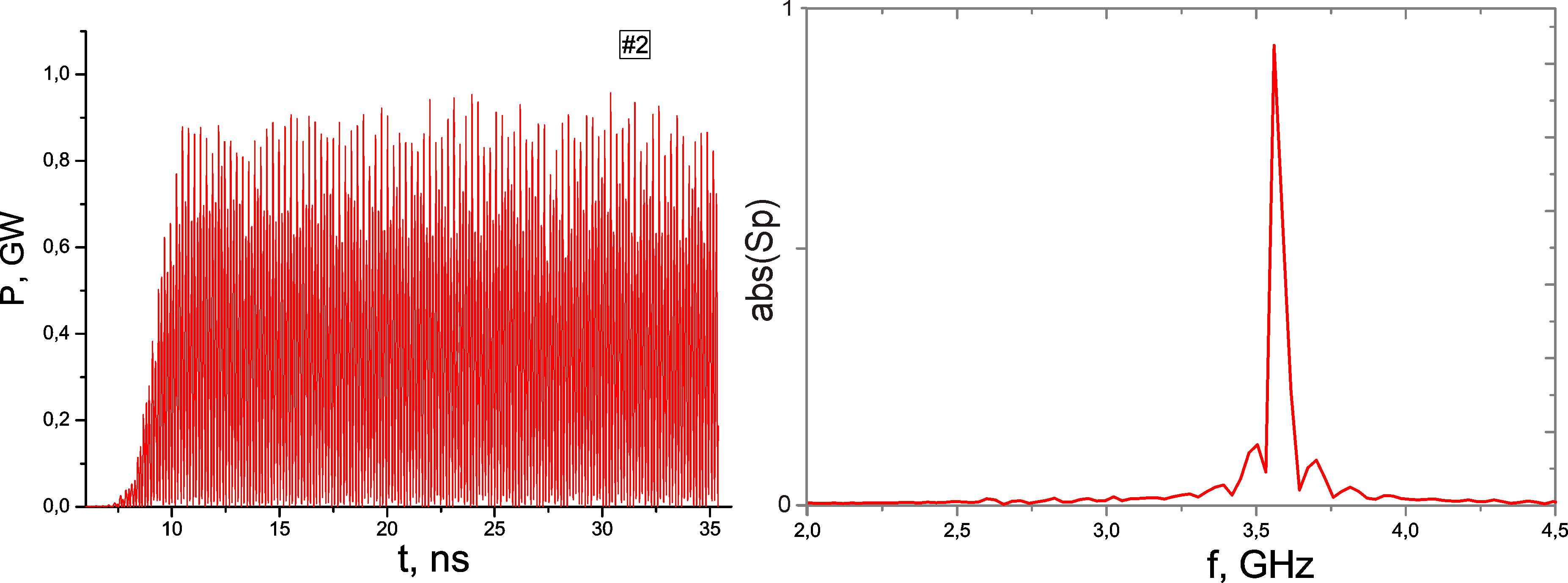}
              \caption{Output power and radiation spectrum of the axial vircator with three-cavity resonator   $\bf{\sharp}$2 for a solid cathode of radius $r_b$=35.5 mm and the cathode-anode gap of 14 mm.}
    \end{figure}

   \noindent

  \begin{figure}[h]
       \centering
        \includegraphics[width=0.8\linewidth]{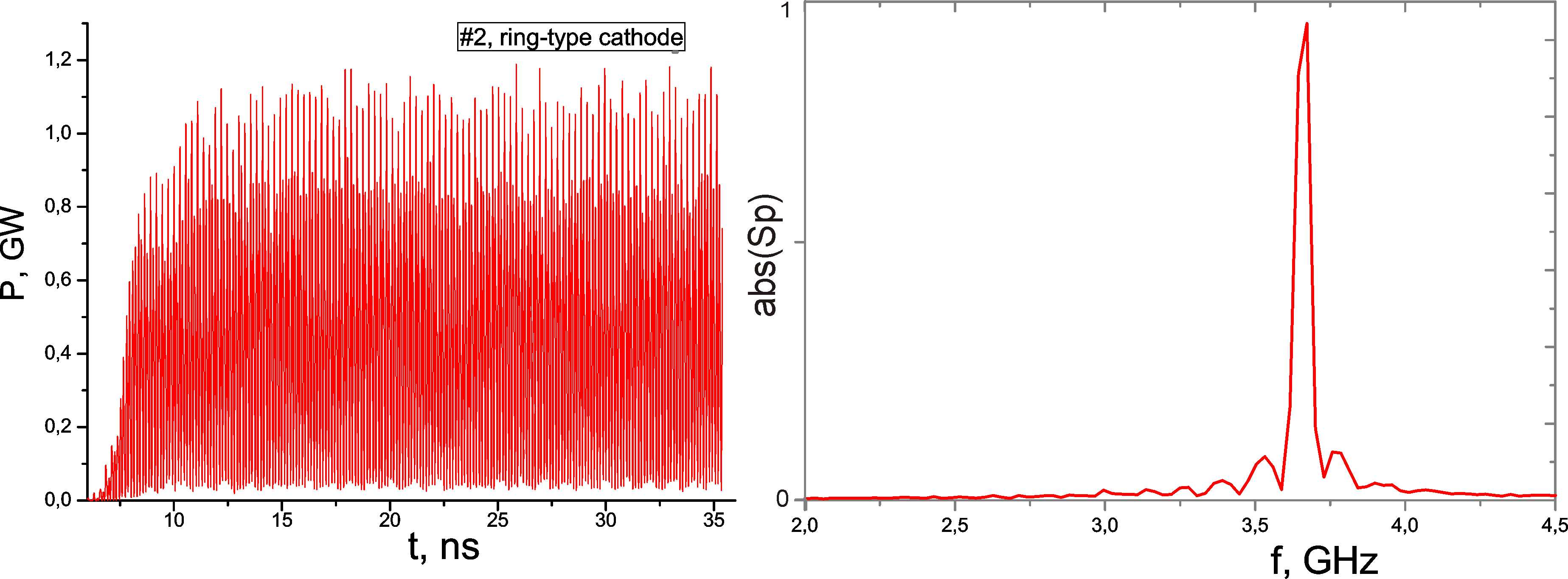}
              \caption{Output power (left) and radiation spectrum (right) in the axial vircator with three-cavity resonator $\bf{\sharp}$2 for a ring-type cathode with inner and outer radii of $r_{b1}$=13.5mm and  $r_{b2}$=35.5mm, respectively, and the cathode-anode gap equal to 13 mm.}
    \end{figure}

   \noindent

  \begin{figure}[h]
       \centering
        \includegraphics[width=0.9\linewidth]{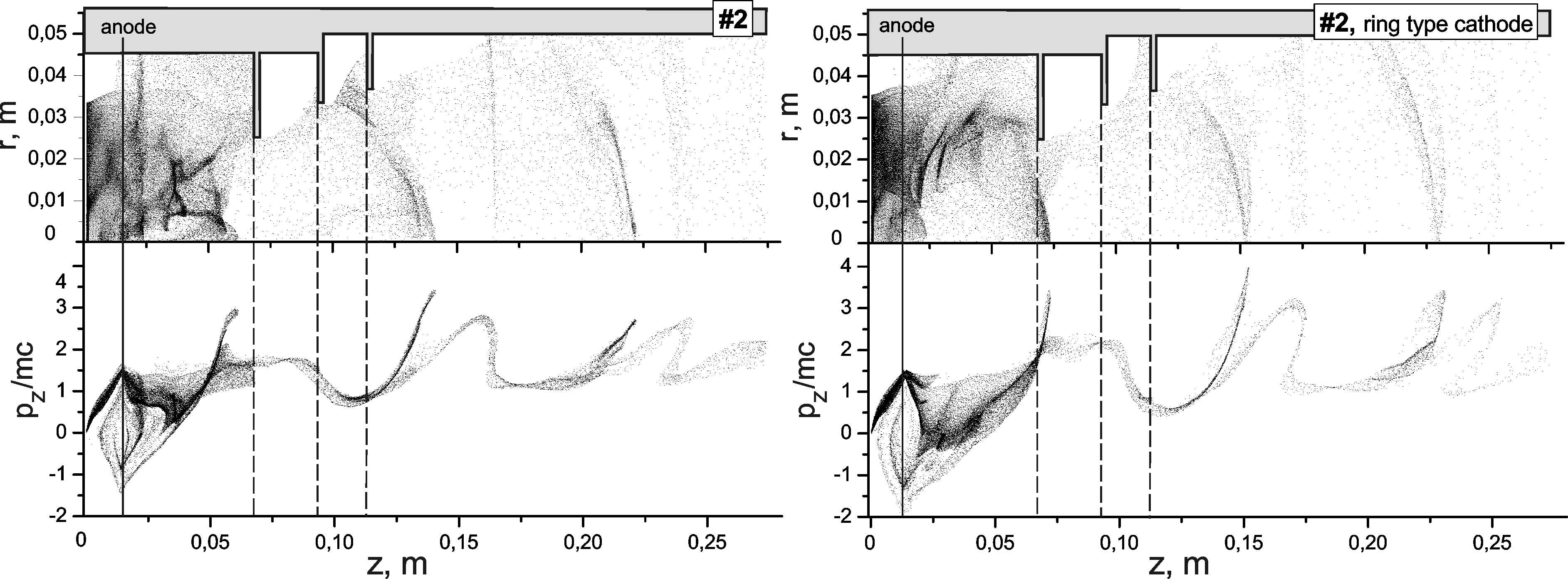}
              \caption{Configuration (top) and longitudinal (bottom) phase portraits of the beam in axial vircator with  three-cavity resonator $\bf{\sharp}$2: left -- for a solid cathode of radius $r_{b}$=35.5 and cathode-anode gap of 14 mm; right -- for a ring-type cathode with inner and outer radii of $r_{b1}$=13.5mm and  $r_{b2}$=35.5mm, respectively, and the cathode-anode gap of 13 mm.}
    \end{figure}

   \noindent

  \begin{figure}[h]
       \centering
        \includegraphics[width=0.8\linewidth]{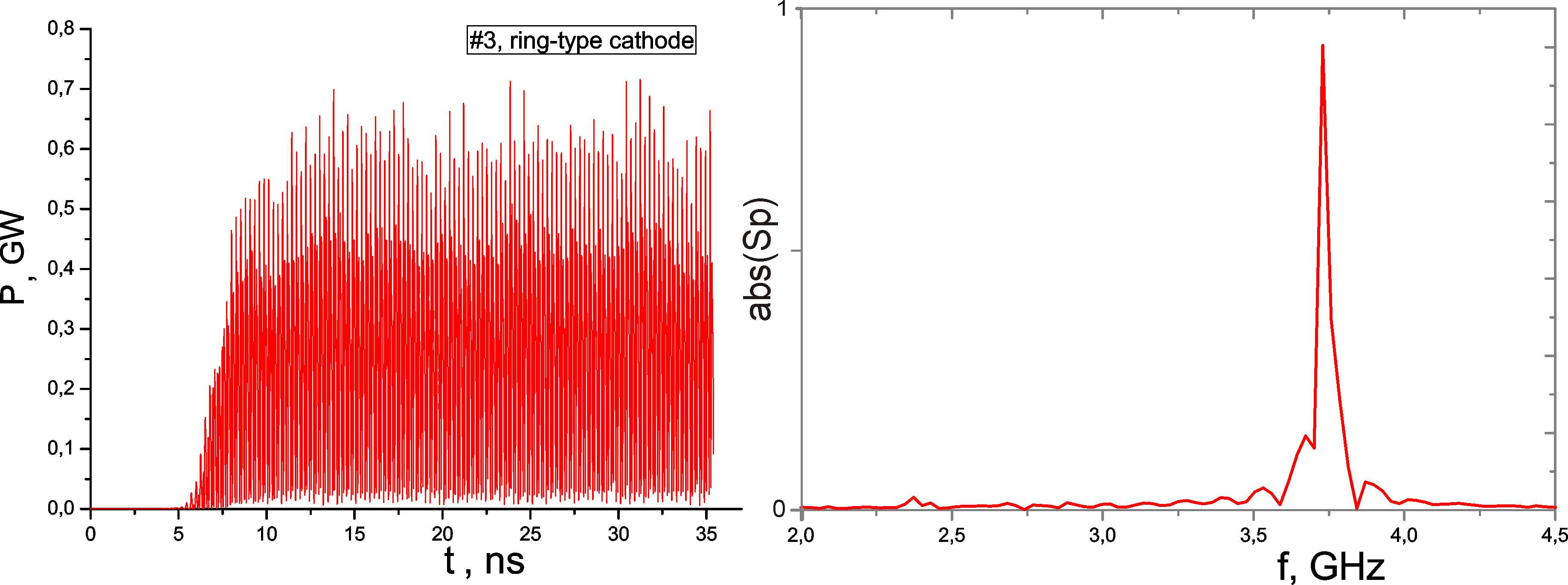}
              \caption{Output power  (left) and radiation spectrum (right) in the axial vircator with three-cavity resonator $\bf{\sharp}$3 and ring-type cathode.}
    \end{figure}

   \noindent

   \begin{figure}[h]
       \centering
        \includegraphics[width=0.6\linewidth]{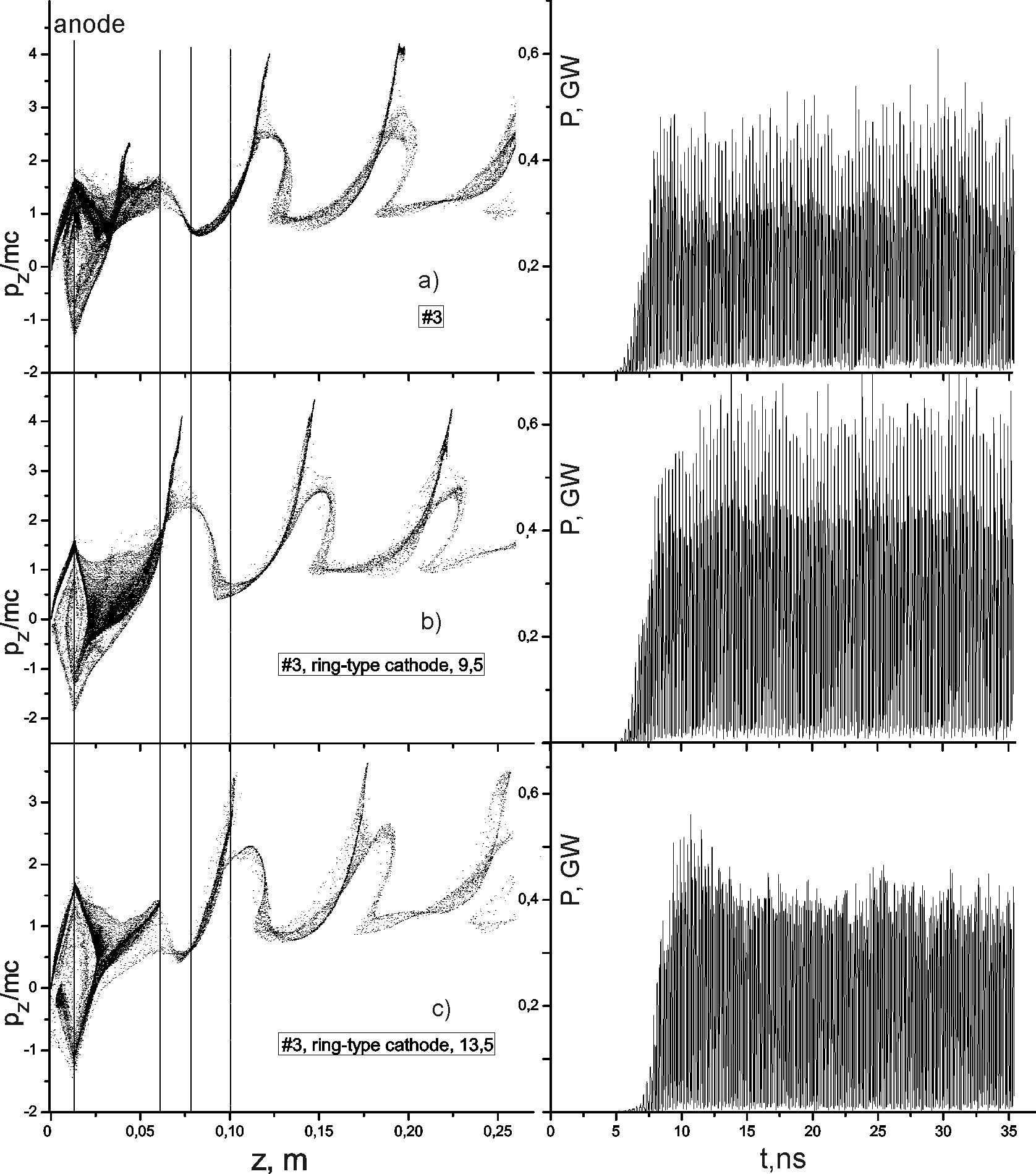}
              \caption{ Longitudinal phase portraits of the beam with three-cavity resonator  $\bf{\sharp}$3 with the cathode-anode gap of 13 mm: (a) -- for a solid cathode of radius $r_{b}$=35.5 mm;  (b) -- for a ring-type cathode with inner and outer radii $r_{b1}$ = 9.5 mm  (the radius of the hole in the cathode) and $r_{b2}$=35.5 mm; (c) -- for a ring-type cathode with inner and outer radii $r_{b1}$=13.5 mm and $r_{b2}$=35.5 mm, respectively.}
    \end{figure}

   \noindent

In this section, we shall consider several alternative geometries of a multicavity resonator for an axial vircator that provide the efficiency greater than 5$\%$ for generation frequency from 3 to 4 GHz at cathode voltage less than 600 kV. Based on the results in \cite{1}, we developed three alternate designs of three-cavity resonators; their dimensions are listed in Table 2.

The operation of each design was simulated with INPIC for a constant cathode voltage of 450 kV. The cathode radius and configuration (solid or ring-type), as well as the cathode-anode gap, are determined so as to provide stable generation and the highest possible values of the output power. In modeling resonator $\bf{\sharp}$1 with INPIC, we have taken the radius of the solid cathode as $r_b$=33.5 mm and the cathode-anode gap as 14 mm. The output power and the radiation spectrum obtained at cathode voltage of 450 kV are shown in Fig. 10. The instantaneous peak power of radiation at a frequency of 3.2 GHz reaches 700 MW. As is seen from the configuration and phase portraits of the beam (Fig. 11), there is a significant number of high-energy particles in the third and output sections of the resonator, which is different from the case of the resonator discussed earlier (see Fig. 8).

\medskip
\mbox{Table2. Dimensions of three-cavity resonators in mm (Fig.1)}
\medskip

\begin{tabular}{|c|c|c|c|c|c|c|c|c|c|c|}
  \hline

 & $L_1$ & $L_2$ & $L_3$ & $L_w$ & $r_1$ & $r_2$ & $r_3$ & $r_4$ & $R$ & $t$ \\
  \hline
  $\bf{\sharp}1$ & 60 & 25 & 17 & 160 & 50 & 25 & 40 & 60 & 70 & 5 \\
 $\bf{\sharp}2$ & 54 & 26 & 20 & 160 & 45 & 25 & 33 & 37 & 50 & 5 \\
  $\bf{\sharp}3$ & 48 & 22 & 17 & 160 & 45 & 21 & 40 & 53 & 62 & 4 \\
  \hline
 \end{tabular}
 \bigskip

For resonator $\bf{\sharp}$2, stable generation with instantaneous peak power of about 800 MW at a frequency of 3.8 GHz was obtained for a solid cathode of radius $r_b$=35.5 mm and a cathode-anode gap of 14 mm. The output power and radiation spectrum at cathode voltage of 450 kV are given in Fig. 12. Replacing the solid cathode by a ring-type one with the inner and outer radii of $r_{b1}$=13.5mm and  $r_{b2}$=35.5mm, respectively, and choosing the cathode-anode gap equal to 13 mm results in about a 20$\%$ increase in the instantaneous peak power at the same frequency. The radiation power and spectrum obtained at cathode voltage of 450 kV are given in Fig. 13. The increase in the output power observed when using a ring-type cathode can possibly be explained by the obvious difference in the phase portraits of the beam (see Fig. 14): the number of high-energy (accelerated by the field) particles in the three-cavity resonator is greater than that obtained for a solid cathode.

The output power and the radiation spectrum simulated with INPIC for the axial vircator with resonator $\bf{\sharp}$3, ring-type cathode ($r_{b1}$=9.5mm and $r_{b2}$=35.5mm), and the cathode-anode gap of 13 mm at cathode voltage of 450 kV are shown in Fig. 15. The basis for selection of the ring-type cathode rests on the analysis of the phase portraits of the beam for solid and ring-type cathodes with different values of the inner diameter $r_{b1}$ (Fig. 16). Analyzing Fig. 16, one can conclude that conversion of the beam energy to HPM is most efficient in the axial vircator with resonator $\bf{\sharp}$3 and a cathode with inner and outer radii $r_{b1}$=9.5 mm and $r_{b2}$=35.5 mm, respectively (Fig. 16b). In this case, most of the particles oscillate near the virtual cathode, and their longitudinal momenta $p_z/mc$ are close to or slightly less than zero; a fraction of high-energy particles is quite effectively "cut off" by the first annulus, though some of  them still enter into the second cavity (section). It is seen from the phase portrait in Fig.16a for the axial vircator with a solid cathode that in the first cavity, the longitudinal momenta $p_z/mc$  of a large number  of particles are close to unity, but not less than zero as in the axial vircator with cathode of  $r_{b1}$= 9.5 mm (Fig. 16b), which indicates a smaller efficiency of the  virtual cathode. One can also see that the particle density  near the virtual cathode for a $r_{b1}$=13.5 mm (Fig. 16c) cathode is rather small, and the output power is not high  in spite of the fact that the first and second annuli effectively cut off high-energy particles (this is because the cathode configuration is not optimal).

The above analysis implies that the effective operation of a three-cavity axial vircator depends strongly on both the resonator geometry, selected so as to effectively cut off the fraction of high-energy particles, and the cathode parameters.

\section{Conclusion}

It is worth noting that the analysis given here  has been made neglecting many aspects that could cause  a noticeable decrease in the generation efficiency, and shortening of the pulse length. Particularly, the cathode plasma expansion \cite{7} and the influence of the energy (velocity) spread of the beam electrons, which necessarily exists in the system, leads to an appreciable lowering of efficiency penalty at practically unchanged radiation spectrum \cite{8} for XOOPIC simulations (in INPIC, this spread is inherently included).

It should also be taken into account in planning the experiments that XOOPIC simulation is performed at constant beam energy and current, set as parameters that could cause
radiation power overestimation \cite{9}, while INPIC simulation makes it possible to easily include the change in the cathode voltage, as well as to simulate the actual current.

Stable generation and high values of instantaneous peak power obtained through using the simulation procedures described in this paper is a good reason for conducting experimental investigation of the suggested three-cavity resonators.

\bigskip

\section{Acknowledgements} We would like to thank Prof. V.G. Baryshevsky and A.A. Gurinovich for cooperation and support given to this work .


\begin{thebibliography}{14}

\bibitem{1} Li Z.-Q., Zhong H.-H., Fan Y.-W., Shu T., Yang J.-H., Yuan C.-W., Xu L.-R., Zhao Y.-S.,
\textit{Chin. Phys. Lett.}, 2008, Vol. 25, N 7, P. 2566;

\bibitem{2} Tikhomirov V.V., Siahlo S.E. [Online] LANL Arxive acc-ph/1309.6486; (Available:
arxiv.org/abs/1309.6486v1)

\bibitem{3} Goplen B., Ludeking L, Smithe D., Warren G. \textit{Comp .Phys.
Comm}. 1995, Vol. 87, P. 54-86;

\bibitem{4} Verboncoeur J. P., Langdon A. B., Gladd N.T.  \textit{Comp .Phys.
Comm}. 1995, Vol. 87, N 1, P. 199-211;

\bibitem{5} Molchanov P.V. \textit{Proc of XXI Int. Seminar Nonlinear Phenomena in Complex Systems Minsk, Belarus, May 20-23}, 2014, to be published;

\bibitem{6} Gurnevich E.A. \textit{Proc of XXI Int. Seminar Nonlinear Phenomena in Complex Systems Minsk, Belarus, May 20-23}, 2014, to be published;

\bibitem{7} Anishchenko S.V., Gurinovich A.A., \textit{J. of Phys.: Conf. Ser.}, 2014, Vol. 490, P. 012116;

\bibitem{8} Gurnevich E.A., Molchanov P.V. [Online] LANL Arxive acc-ph/1407.8441; (Available:
arxiv.org/abs/1407.8441)

\bibitem{9} Shlapakovski A.S., Kweller T., Hadas Y et.al.\textit{IEEE Trans. on Plasma Sci.}, 2009, Vol. 37, N 7, P. 1233-1241;



\end{thebibliography}
\end{document}